\documentclass[twocolumn,aps,letter,showpacs]{revtex4}
\usepackage[T1]{fontenc}
\usepackage[cp852]{inputenc}
\usepackage[a4paper]{geometry}
\usepackage{amsmath}
\usepackage{amssymb}
\usepackage{graphicx}
\usepackage{esint}

\makeatletter

\providecommand{\tabularnewline}{\\}

\@ifundefined{textcolor}{}
{%
 \definecolor{BLACK}{gray}{0}
 \definecolor{WHITE}{gray}{1}
 \definecolor{RED}{rgb}{1,0,0}
 \definecolor{GREEN}{rgb}{0,1,0}
 \definecolor{BLUE}{rgb}{0,0,1}
 \definecolor{CYAN}{cmyk}{1,0,0,0}
 \definecolor{MAGENTA}{cmyk}{0,1,0,0}
 \definecolor{YELLOW}{cmyk}{0,0,1,0}
}

\usepackage{dcolumn}\usepackage{bm}\usepackage{float}\usepackage{units}\usepackage[all]{xy}\usepackage{multirow}\usepackage{array}\usepackage[titletoc]{appendix}\usepackage{rotating}

\makeatother

\begin{document}
\global\long\def\bea{ 
\begin{eqnarray}
\end{eqnarray}
 }
 \global\long\def\eea{{eqnarray}}
 \global\long\def\bit{\begin{itemize}\end{itemize}}
 \global\long\def\eit{{itemize}}
 \global\long\def\be{ 
\begin{equation}
\end{equation}
 }
 \global\long\def\ee{{equation}}
 \global\long\def\ra{\rangle}
 \global\long\def\la{\langle}
 \global\long\def\U{\widetilde{U}}


\global\long\def\bra#1{{\langle#1|}}
 \global\long\def\ket#1{{|#1\rangle}}
 \global\long\def\bracket#1#2{{\langle#1|#2\rangle}}
 \global\long\def\inner#1#2{{\langle#1|#2\rangle}}
 \global\long\def\expect#1{{\langle#1\rangle}}
 \global\long\def\e{{\rm e}}
 \global\long\def\proj{{\hat{{\cal P}}}}
 \global\long\def\tr{{\rm Tr}}
 \global\long\def\H{{\hat{H}}}
 \global\long\def\Hdag{{\hat{H}}^{\dagger}}
 \global\long\def\Lop{{\cal L}}
 \global\long\def\Ehat{{\hat{E}}}
 \global\long\def\Edag{{\hat{E}}^{\dagger}}
 \global\long\def\Shat{\hat{S}}
 \global\long\def\Sdag{{\hat{S}}^{\dagger}}
 \global\long\def\Ahat{{\hat{A}}}
 \global\long\def\Adag{{\hat{A}}^{\dagger}}
 \global\long\def\U{{\hat{U}}}
 \global\long\def\Udag{{\hat{U}}^{\dagger}}
 \global\long\def\Zhat{{\hat{Z}}}
 \global\long\def\Phat{{\hat{P}}}
 \global\long\def\Op{{\hat{O}}}
 \global\long\def\id{{\hat{I}}}
 \global\long\def\x{{\hat{x}}}
 \global\long\def\P{{\hat{P}}}
 \global\long\def\Px{\proj_{x}}
 \global\long\def\Pr{\proj_{R}}
 \global\long\def\Pl{\proj_{L}}

\newcommand{\angstrom}{\text{\normalfont\AA}}

\title{Bohr's correspondence principle for atomic transport calculations}
\author{Viviana P. Ramunni} \author{Alejandro M.F. Rivas}
\affiliation{Conicet, Av. Rivadavia 1917,
 (C1033AAJ) Buenos Aires, Argentina.}
\date{\today}
\begin{abstract}
In this work we perform a comparison between Classical Molecular Static (CMS) and
quantum Density Functional Theory (DFT) calculations in
order to obtain 
the 
diffusion coefficients for diluted \emph{Fe-Cr} alloys.
We show that, in accordance with Bohr's correspondence principle, as the 
size of the atomic cell (total number of atoms) is increased, quantum results
with DFT approach to the classical ones obtained with CMS. 
Quantum coherence effects 
play a crucial role in the  difference 
arising between CMS and  DFT calculations.
Also, thermal contact with the environment destroys quantum coherent effects
making the classical behavior to emerge.
Indeed, CMS calculations
are in good agreement with available experimental data.
We claim that, the atomic diffusion process in metals is a classical 
phenomena.
Then, if reliable semi empirical potentials are available,
a classical treatment of the atomic transport in metals is much convenient than DFT.


\end{abstract}
\maketitle
To characterize the crossover between the quantum and the classical
worlds is a fundamental quest of modern physics. For some systems, it is clear
that, classical physics arises from quantum physics in the large-number limit.
This is the Bohr's correspondence principle
\cite{bohr}.

It is not fully understood yet how the many-particle limit gives
rise to classical physics, and how much of quantum physics still remains.
In this framework, the environment and its temperature plays an important
role for the decoherence process \cite{zurek}. 
However, quantum coherence effects at ambient temperatures
where observed in biological systems \cite{quanphot}.

Also, for many body systems we must deal with both electrons and atomic
nuclei dynamics. Due to their masses the nuclei move much slower than
the electrons. Then in the Born\textendash{}Oppenheimer approximation
\cite{BORNOPP} the nuclei generate a static
external potential in which the electrons are moving. 

First principles ({\it Ab initio}) quantum mechanical method,
such as DFT, are employed to obtain the electronic behavior.
DFT reduces the quantum many-body
problem to the use of functionals of the electron density \cite{DFT}, it is 
presently the most successful approach to compute the electronic
structure of matter. Its applicability ranges from atoms, molecules
and solids to nuclei and liquids \cite{DFT}. 
The electron density determines the potential energy surface
that represents the force field where the nuclei dynamics occurs.

In a different approach, a CMS treatment, employs phenomenological 
semi-empirical potentials in order to estimate
this force field. Then, the quantum nature of electronic structure is not taken into
account. There is a wide variety of semi-empirical potentials,
which vary according to the atoms being modeled.
\emph{Ab Initio} simulations take into account the quantum 
nature of the electrons, which implies in a higher computational cost
than CMS. 
Hence ab initio simulations are limited to smaller systems.

In this work atomic diffusion, in \emph{Fe-Cr} diluted alloys is
studied with both CMS and quantum DFT calculations, in the context of
a multi-frequency model. We show that,  
in accordance with Bohr's correspondence principle,
as the total number of atoms is increased, quantum results
with DFT recover the classical ones obtained with CMS.

Quantum coherence effects, which are only taken into account by DFT,
play a crucial role for both the convergence issue of the DFT
results as a function of supercell size, as well as, in the difference 
arising between classical CMS and quantum DFT calculations.

In addition, CMS calculations are in good agreement with available experimental 
data for both solute and solvent diffusion coefficients.
This may not be surprising for a macroscopic system especially for high temperatures 
that destroys any quantum coherent effect.

Diffusion plays an important role in the kinetics of many materials
processes. Experimental measurements of diffusion coefficients are
expensive, difficult and in some cases nearly impossible. A complimentary
approach is to determine diffusivities in materials by atomistic computer
simulations. In addition to predicting diffusion coefficients, computer
simulations can provide insights into atomic mechanisms of diffusion
processes, creating a fundamental framework for materials design strategies.

Also, \emph{Fe-Cr} alloys at low temperatures has important technological
consequences. Due to their good resistance to void swelling \cite{feCRcentral1,fecrcentral2},
\emph{Fe-Cr} based alloys are of special interest for nuclear applications
(in Generation IV and fusion reactors). 

Atomic transport theory allows to express the diffusion coefficients
in terms of the atomic frequency jumps, this is commonly known as
the multi-frequency model \cite{LEC78}.
Recently, attends were made in order to describe the diffusion process
by obtaining numerically the needed frequency jumps with DFT calculations.
Although disagreement between the experimental and \emph{ad-initio}
based calculated diffusion coefficients where observed in bcc alloys
such as \emph{Ni-Cr} and \emph{Ni-Fe} \cite{TUC10}, and for $\alpha$\emph{Fe-Ni}
and $\alpha$\emph{Fe-Cr} alloys \cite{CHO11} as well as for $Mg$,
$Si$ and $Cu$ diluted in fcc $Al$ \cite{MAN09}. However, in a
recent work Huang et \textit{al.} \cite{ACTA10},  for
\emph{Fe} based diluted alloys, have performed DFT based calculations
for the tracer diffusion coefficient with a larger number of atoms
($128$ instead of $54$ as in \cite{CHO11}) that are in good agreement
with the experimental data.

On the other hand, one of us has recently shown \cite{RAM13}, that
tracer diffusion coefficients performed with CMS based calculations
in diluted \emph{Ni-Al} and \emph{Al-U} fcc alloys are in excellent
agreement with available experimental data for both systems.

\begin{figure}[h]
\begin{centering}
\hspace{-28.5pt}\includegraphics[width=4.8cm]{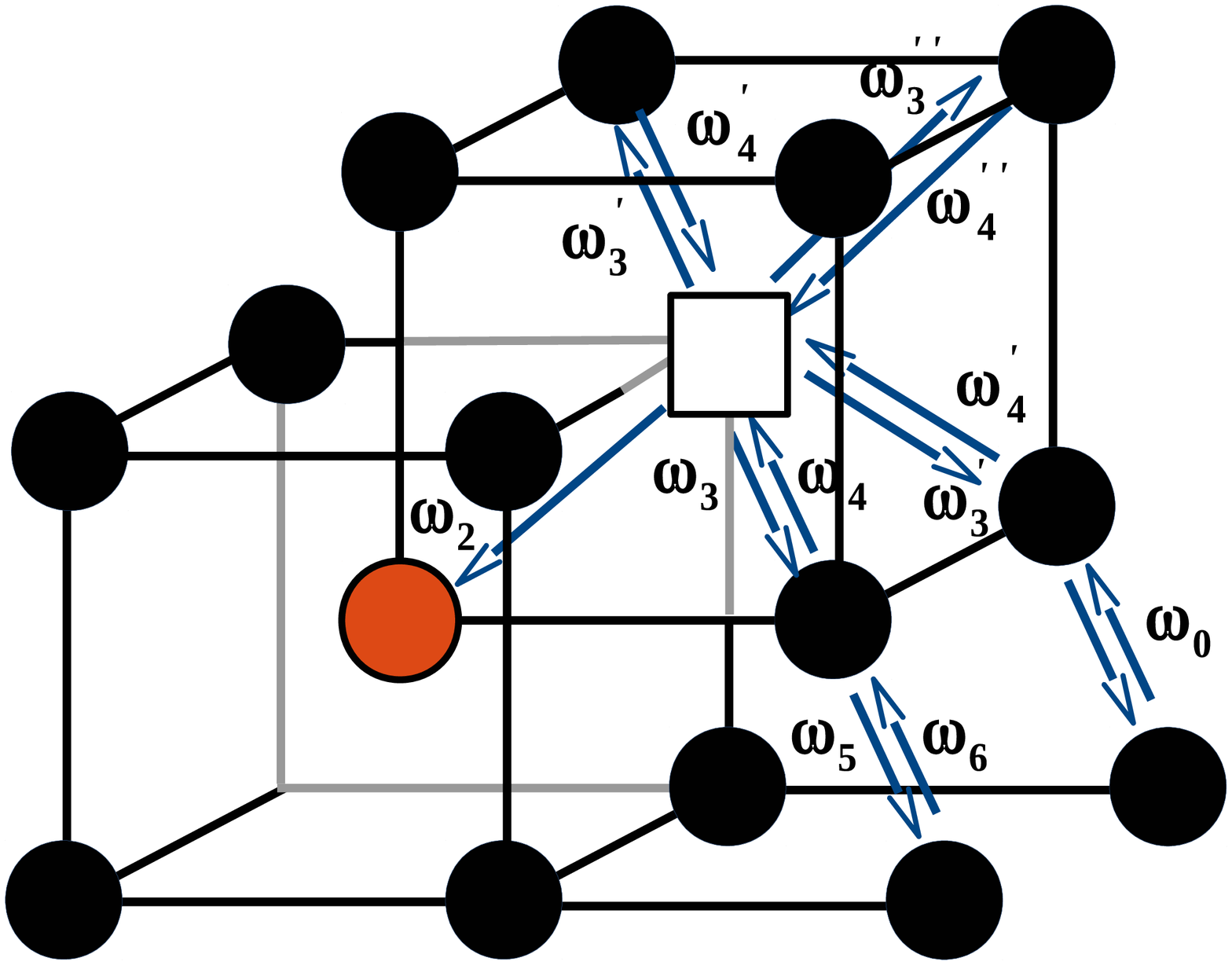}
\includegraphics[bb=130bp 155bp 792bp 719bp,clip,width=4.8cm]{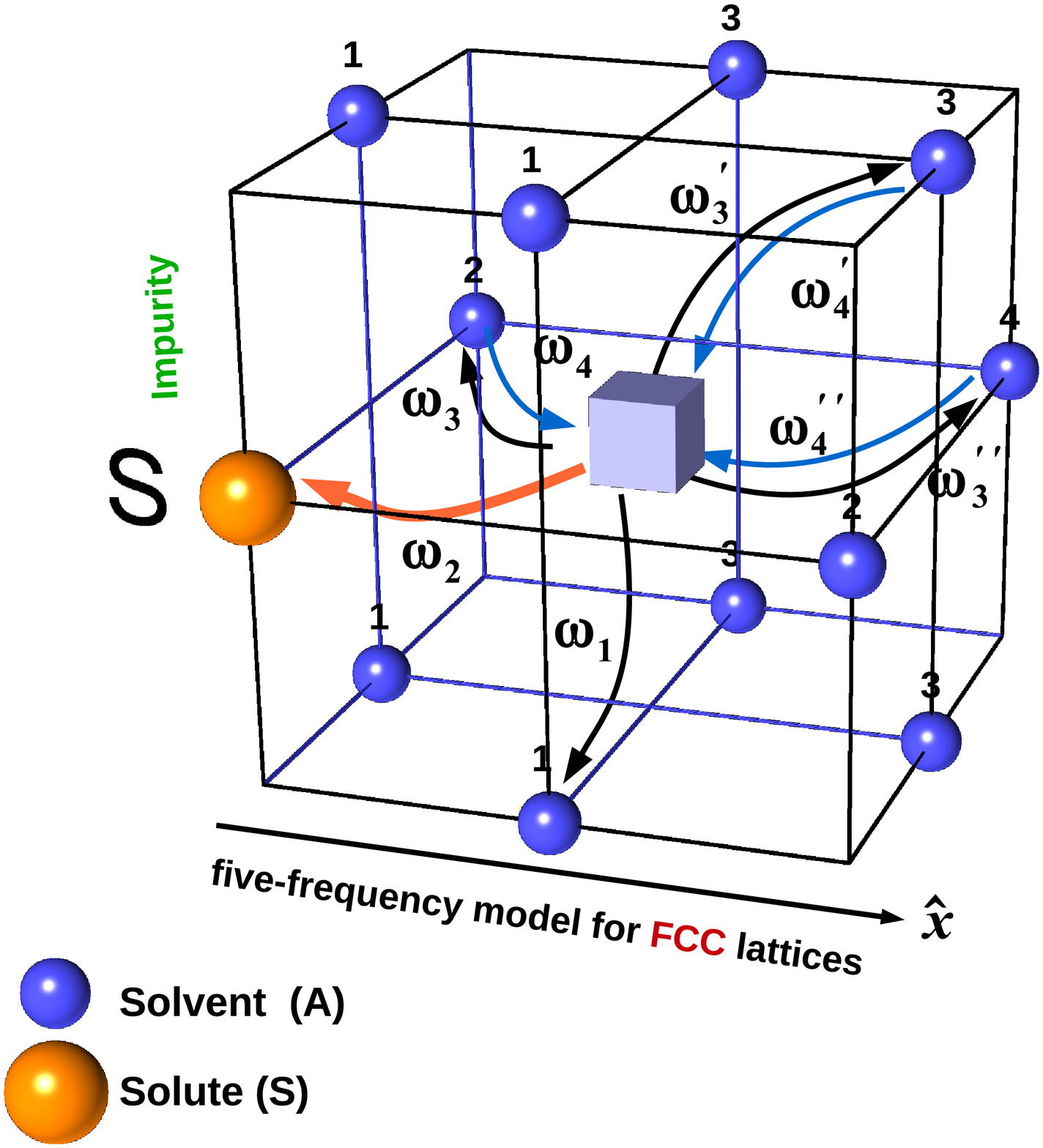}
\caption{(Color online) The frequencies involved in the second binding model for bcc and fcc
lattices. In black/orange circles, respectively are represented the
solvent and solute atoms.}
\label{6f-V} 
\par\end{centering}
\end{figure}

We focus here on the tracer self- and solute diffusion coefficients in a
binary \emph{A-S} alloy in the diluted limit. For diffusion mediated
by vacancies analytical expressions, in terms of the frequency
jumps, where calculated by Allnatt \cite{ALL81a} and Le Claire
\cite{LEC78} for fcc and bcc lattices respectively. In the $2nd$-nearest-neighbor
binding model, we identify the jumps as in Fig. \ref{6f-V}.
The self-diffusion coefficient can be written as,
\begin{equation}
D_{A}^{\star}=a^{2}\omega_{0}C_{V}f_{0},\label{DA00-1}
\end{equation}
where $a$ is the lattice parameter, $\omega_{0}$ is the atom-vacancy
exchange frequency in pure $A$ and $f_{0}$, the self-diffusion correlation
factor that is $f_{0}=0.7272$ \cite{LEC78} or $f_{0}=0.7814$  \cite{ALL81a},
for bcc or fcc metals respectively. At thermodynamic equilibrium the vacancy concentration
depends on the temperature $T$ as, 
\begin{equation}
C_{V}=\exp\left(-\frac{E_{f}^{V}-TS_{f}^{V}}{k_{B}T}\right)\label{cv0}
\end{equation}
 where $k_{B}$ is the Boltzmann constant while $E_{f}^{V}$ and $S_{f}^{V}$
respectively denote the energy and entropy formation of the vacancy
in pure $A$.

The impurity diffusion coefficient, $D_{S}^{\star}$, depends on
several jump frequencies, corresponding to the exchanges of the vacancy
with the solute atom $S$ and with the solvent atoms $A$ near $S$, see Fig. \ref{6f-V}.
\begin{equation}
D_{S}^{\star}=a^{2}\omega_{2}f_{S}C_{V}\left(\frac{\omega_{4}}{\omega_{3}}\right)\,,\label{DBB2}
\end{equation}
where $\omega_{2}$ is the $S$-vacancy exchange frequency
and $f_{S}$ is the solute correlation factor. 

For bcc lattices, in the formalism of Le Claire\cite{LEC78}, the correlation factor 
is
\begin{equation}
f_{S}^{bcc}=\frac{1-t}{1+t} ,\label{fS-BCC}
\end{equation}
where $t$ is expressed in terms of the jump frequencies as:
\begin{equation}
\!t\!=\!\frac{-\omega_{2}}{\omega_{2}\!+\!3\omega_{3}\!+\!3\omega^{\prime}_{3}\!+\!3\omega^{\prime\prime}_{3} \!-\!\frac{\omega_{3}\omega_{4}}{\omega_{4}+F\omega_{5}} 
\!-\!\frac{2\omega^{\prime}_{3}\omega^{\prime}_{4}}{\omega^{\prime}_{4}+3F\omega_{0}} 
\!-\!\frac{\omega^{\prime\prime}_{3}\omega^{\prime\prime}_{4}}{\omega^{\prime\prime}_{4}+7F\omega_{0}} 
} ,\label{t-BCC}
\end{equation}
with $F=0.512$ in (\ref{t-BCC}).
 %
For fcc lattices, the solute correlation factor \cite{ALL81a} is
\begin{equation}
f_{S}^{fcc}=\left\{ \frac{2\omega_{1}+7\omega_{3}F^{fcc}}{2(\omega_{1}+\omega_{2})+7\omega_{3}F^{fcc}}\right\} ,\label{fS-FCC}
\end{equation}
with $F^{fcc}$ expressed as a function of $u=\omega_{4}/\omega_{0}$ as
\begin{equation}
7(1-F^{fcc})=\frac{u(\xi_{1}u^{3}+\xi_{2}u^{2}+\xi_{3}u+\xi_{4})}{\xi_{5}u^{4}+\xi_{6}u^{3}+\xi_{7}u^{2}+\xi_{8}u+\xi_{9}},\label{F-FCC}
\end{equation}
whith the $\zeta_{i}$ coefficients calculated by Koiwa in \cite{KOI83}.

According to the transition-state theory, in a system of $N$ atoms,
the exchange frequency between a vacancy and an atom is,
\begin{equation}
\omega_{i}=\nu_{0}\exp\left(-\frac{G_{m}^{i}}{k_{B}T}\right)=\nu_{0}\exp\left(\frac{TS_{m}-H_{m}^{i}}{k_{B}T}\right). \label{OMEGAi}
\end{equation}
In (\ref{OMEGAi}), $G_{m}^{i}$ is the migration Gibbs free energy and the pre-exponential
term, the "attempt frequency" $\nu_{0}$, is of the order
of the Debye frequency. The Gibbs free energy is given
by $G_{m}=H_{m}-TS_{m}$, where $S_{m}$ is the migration entropy,
while $H_{m}$ is the enthalpy. As the volume is kept constant and
the pressure is considered null, $H_{m}=E_{m},$ where $E_{m}$
is the internal migration energy. Hence, following Vineyard's formulation \cite{VIN57},
the migration frequency jumps are given by
\begin{equation}
\omega_{i}=\nu_{0}^{\star}\exp(-E_{m}^{i}/k_{B}T).\label{eq:nuE}
\end{equation}
In (\ref{eq:nuE}), $E_{m}^{i}$ are the vacancy migration energies at $T=0K$, while 
\begin{equation}
\nu_{0}^{\star}\!=\!\left({\displaystyle \prod_{i=1}^{3N-3}\nu_{i}^{I}}\right)/\left({\displaystyle 
\prod_{i=1}^{3N-4}\nu_{i}^{S}}\right)\!=\!\nu_{0}\exp\left(\frac{S_{m}}{k_{B}}\right)\label{PInu},
\end{equation}
with $\nu_{i}^{I}$ and $\nu_{i}^{S}$ the frequencies of the normal
vibrational modes at the initial and saddle points, respectively. 

We present our numerical results applied to \emph{Fe-Cr} diluted
alloys. Above the melting temperature $T_{\alpha\gamma}=1183K$, \emph{Fe-Cr} alloys
develop a paramagnetic fcc phase, while for lower temperature the structure is bcc. In this bcc phase,
a magnetic transitions occurs
from ferromagnetic, below the Curie temperature $T_{C}=1043K$, to paramagnetic states.

In the case of the bcc phase, we performed both DFT and CMS calculations.
For DFT calculations, we have employed localized
basis sets as implemented in SIESTA code \cite{SIESTA}. We have also
considered spin polarization and GGA approximation in all calculations.
Core electrons are replaced by nonlocal norm-conserving pseudo potentials
as in Ref. \cite{MAR12}. Valence electrons are described by linear
combinations of localized pseudoatomic orbitals. The basis sets for
both elements consist in two and three localized functions for the
4s and 4p states, respectively, and five for the the 3d states. The
maximum cutoff radius is $5.1$ \AA{}. Calculations were carried
out with $54$ and $128$ atom supercells, using respectively a $7\times7\times7$
and $4\times4\times4$ k-point grid, and the Methfessel-Paxton broadening
scheme with a $0.3eV$ width. The migration barriers have been determined
using SIESTA coupled to the Monomer \cite{RAM06}. 

In CMS calculations the atomic interaction are represented by EAM
potentials. For the \emph{Fe-Cr} system in the bcc lattice we have used the potential developed
by Mendelev \emph{et al.} \cite{MEN03}, for the pure elements \emph{$Fe$}
and $Cr$, as well as, for the cross \emph{Fe-Cr} term. 
While for the hight temperature fcc phase, where only CMS calculations were performed, 
we have used the potential developed by Bonny \emph{et al.} \cite{BON13}.
For all classical calculations we use a christallyte of $8\times8\times8$
with periodic boundary conditions, that is $1024$ and $2048$ atoms
for bcc and fcc respectively. 
We have verified, for the bcc phase, that 
the results do not change if we employ a christallyte of $128$ atoms.
We obtain the equilibrium positions
of the atoms by relaxing the structure via the conjugate gradients
technique. The lattice parameters that minimize the crystal structure
energy are $a_{Fe}=2.866$\AA{}, and $a_{Fe}=3.562$\AA{}, for bcc
and fcc structures, respectively. 

In Table \ref{T4} we show our calculations of the activation energies, formation and migration, in a perfect bcc $Fe$ lattice. We show both, DTF
(with $54$ and $128$ atoms) together with CMS calculations. Initial
and saddle points configurations and their respective energies are calculated with
the Monomer method \cite{RAM06}.
\begin{table}[ht]
\protect
\begin{centering}
\caption{Energies and lattice parameters for the pure bcc $Fe$ lattice obtained by DFT calculations with $54$ and $128$ atoms and by CMS calculations.}
\label{T4}
\par\end{centering}
\begin{centering}
\protect 
\par\end{centering}
\centering{} %
\begin{tabular}{llccc}
\hline 
 &  & bcc -$Fe$  &  & \tabularnewline
\hline 
\hline 
 & $DFT_{54}$ \,  & \, $DFT_{128}$ \,  & \, $CMS$ \,  & Exp.\tabularnewline
\hline 
\,$a$(\AA{}) \,  & \, $2.885$\,  & \, $2.885$\,  & \, $2.866$ \,  & \, $2.866$ \, \tabularnewline
\,$E_{f}^{V}(eV)$ \,  & \, $2.18$ \, & \, $2.05$ \,  & \,$1.72$ \,  & \,$1.79\pm0.1$ \tabularnewline
\,$E_{m}^{0}(eV)$ \,  & \, $0.67$ \, & \, $0.68$ \,  & \, $0.68$ \,  & \tabularnewline
\,$E_{m}^{2}(eV)$\,  & \, $0.57$ \,  & \, $0.56$ \,  & \, $0.562$ \,  & \tabularnewline
\,$E_{m}^{3}(eV)$\,  & \, $0.67$ \,  & \, $0.67$ \,  & \, $0.67$ \,  & \tabularnewline
\,$E_{m}^{4}(eV)$\,  & \, $0.64$ \,  & \, $0.63$ \,  & \, $0.625$ \,  & \tabularnewline
\,$E_{m}^{3^{\prime}}(eV)$\,  & \, $0.63$ \,  & \, $0.60$ \,  & \, $0.558$ \,  & \tabularnewline
\,$E_{m}^{4^{\prime}}(eV)$\,  & \, $0.61$ \,  & \, $0.60$ \,  & \, $0.599$ \,  & \tabularnewline
\,$E_{m}^{3^{\prime\prime}}(eV)$\,  & \, $0.60$ \,  & \, $0.58$ \,  & \, $0.542$ \,  & \tabularnewline
\,$E_{m}^{4^{\prime\prime}}(eV)$\,  & \, $0.59$ \,  & \, $0.59$ \,  & \, $0.585$ \,  & \tabularnewline
\,$E_{m}^{5}(eV)$\,  & \, $0.64$ \,  & \, $0.63$ \,  & \, $0.627$ \,  & \tabularnewline
\end{tabular}
\end{table}
We can observe from Table \ref{T4} that, as the number of atoms is
increased, the energies calculated with DFT get closer to the classical
ones. This effect is particularly important for the vacancy formation energy where the result obtained with CMS calculation is in accordance with the experimental result measured in \cite{EFV}.

For the fcc paramagnetic phase, occuring above $T_{\alpha\gamma}=1183K$, formation and activation energies from CMS calculations in are displayed in Table \ref{T6}.
\begin{table}[htdp]
\begin{centering}
\caption{Activation energies in paramagnetic fcc \emph{Fe-Cr} from CMS calculations, using
 potential of Ref.\cite{BON13}. }
\label{T6}
\par\end{centering}
\centering{} %
\begin{tabular}{|c|c|c|c|c|c|c|c|c|c|}
\hline 
$E_{f}^{V}$ \,  & \,$E_{m}^{0}$\,  & \,$E_{m}^{1}$\,  & \,$E_{m}^{2}$\,  & \,$E_{m}^{3}$\, 
& \,$E_{m}^{4}$\,  & \,$E_{m}^{3^{\prime}}$\,  & \,$E_{m}^{4^{\prime}}$\,  & \,$E_{m}^{3^{\prime\prime}}$\,  
& \,$E_{m}^{4^{\prime\prime}}$\,\tabularnewline
\hline 
$1.87$ & $0.64$ & $0.66$ & $0.65$ & $0.76$ & $0.72$ & $0,66$ & $0.60$ & $0.70$ & $0.63$\tabularnewline
\hline 
\end{tabular}
\end{table}

In order to compute the jump frequencies, we use expression (\ref{eq:nuE}), with the migration energies $E_{m}^{i}$ reported in Tables \ref{T4} and \ref{T6}, respectively for bcc and fcc phases. For the pre-factor (\ref{PInu}), we have taken the experimental values of the migration entropy $S_{m}=2.1k_{B}$, as
reported in \cite{ENTropmig}, in all cases. While the Debye frequency has been
taken as $\nu_{D}=10^{13}Hz$. 

For the vacancy concentration in (\ref{cv0}), the formation entropy
has been taken as $S_{f}^{V}=4.1k_{B}$, from DFT calculations performed in 
\cite{entropDFT}, while for CMS calculations, we used
$S_{f}^{V}=2.3k_{B}$, as obtained 
in \cite{MEN09}.

Once the jump frequencies in the multi-frequency model have been computed,
the diffusion coefficients are calculated using analytical expressions
(\ref{DA00-1}) and (\ref{DBB2}). Also, it has been observed that, 
due to spontaneous magnetization \cite{IIJ88}, 
the self-diffusion coefficient deviates from a linear Arrhenius relationship, below the Curie temperature. 
This magnetization effects are, as usually, taken into account, as
a correction of the activation energies $Q$ for the ferromagnetic phase, from those in the paramagnetic $Q_{p}$ (in Table \ref{T4}) such that,
\begin{equation}
Q=Q_{p}\left(1+\alpha_{X}s_{X}^{2}(T)\right) ,\label{eq:Fex}
\end{equation}
with $X=Fe$ or $Cr$ and $s_{X}(T)$ is the ratio of
the spontaneous magnetization at $T$ to that at $T=0K$
\cite{CRA71}. While $s_{X}(T)=0$, in the full temperature range of
the paramagnetic phase. In this respect, a direct estimation of these
parameters from first principles would be of great interest. 
Here, as in \cite{LEE90}, we interpolate
the values of $\alpha s^{2}(T)$ in Ref. \cite{CRA71} for both, solute and solvent atoms. 

\begin{figure}[h]
\begin{centering}
\includegraphics[width=9cm,height=8cm,keepaspectratio,angle=-90]{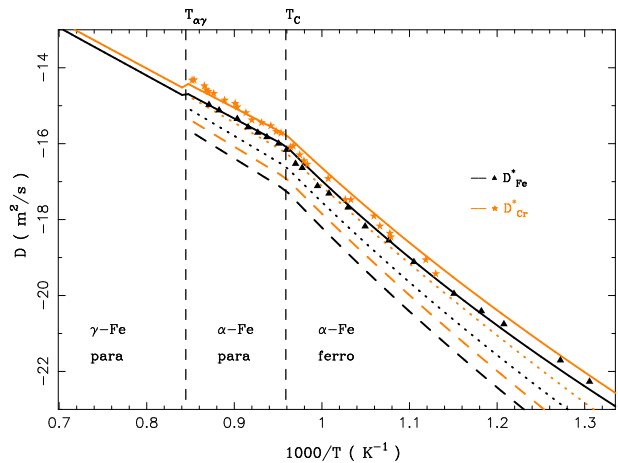}
\vspace{0cm}
 \caption{(Color online) Self-diffusion (in black) and $Cr$ impurity (in orange) diffusion coefficients in $Fe$ from CMS and DFT calculations. Full lines correspond to CMS
calculations while dotted and dashed lines for DFT with $128$ and
\emph{$54$} atoms respectively. Experimental values for $D_{Fe}^{\star}$
and $D_{Cr}^{\star}$, obtained from Refs. \cite{IIJ88} and \cite{LEE90}
are plotted with triangles and stars respectively.}
\label{lupa-1}
\par\end{centering}
\centering{} 
\end{figure}

In Figure \ref{lupa-1}, we show the calculated $D_{Fe}^{\star}$
and $D_{Cr}^{\star}$, using equations (\ref{DA00-1}) and (\ref{DBB2})
respectively, with the activations energies in Tables \ref{T4}
and \ref{T6} for bcc and fcc phases, respectively. As we already mentioned,
for the bcc phase we performed DFT, with $54$ and $128$ atoms, as well as,
CMS calculations. 
Also in figure \ref{lupa-1}, experimental data taken from 
Refs. \cite{IIJ88} and \cite{LEE90} are plotted
with triangles and stars, respectively for $D_{Fe}^{\star}$ and 
$D_{Cr}^{\star}$. 
As can be observer in Fig. \ref{lupa-1}, below the solvent melting
temperature $T_{\alpha\gamma}=1043^{\circ}C$, 
in accordance with Bohr's correspondence principle,
as the size of the atomic cell (total number of atoms) is increased,
quantum results with DFT approach to the classical ones obtained with
CMS.

Note that the bcc supercell with $54$ atoms, corresponding to a christallyte of $3\times3\times3$, 
has a length of $8.59$\AA{}, and for $128$ atoms, the $4\times4\times4$
christallyte has a length of $10.14$\AA{}.
In both cases, this length is lower than typical electronic quantum coherence length 
which are of nanometer order ($4nm$ on $Cu$ \cite{elecoh}). Then, quantum coherence effects, which are only taken into account by DFT,
play a crucial role for both the convergence issue of the DFT
results as a function of supercell size, as well as, in the difference 
arising between classical CMS and quantum DFT calculations.

Instead, with CMS, the semimpirical potential are phenomenological and
the quantum coherence effects are not taken into account. Moreover,  CMS calculations 
with $128$ or $1024$ atoms, give the same results.

In addition, we must not expect to observe quantum effects for such macroscopic systems especially 
at the high temperatures here described. In that case, the
interaction of the system with the thermal environment implies in
decoherence effects, where the classical limit is expected to be recovered
\cite{zurek}. 

Indeed, our results obtained with CMS
calculations are in good agreement with available experimental
data for both, tracer solute and solvent diffusion coefficients.
Note that, similar results have been obtained using, the also classical method, Kinetic Monte Carlo algorithm with temperature dependent pair interactions \cite{MARACTA14}.

It must be emphasized that the agreement between CMS based calculations and experimentally measured diffusion coefficients is not fortuitous,  it has been recently observed for diffusion in \emph{Al-U} and \emph{Ni-Al} fcc lattices \cite{RAM13}. While for this former DFT calculation underestimated the diffusion coefficients \cite{zach}.

Several possible explanations of the fact that DFT calculation are not 
in agreement with experiments for the diffusion coefficients where argued in 
\cite{TUC10}. We claim here that this is due to quantum coherence effects arising
from DFT calculations for the size of the simulation cell being small.
As the size of the simulation cell is increased the DFT results converge to the experimental values that can be obtained with CMS calculations which is much less expensive.

For the fcc phase, where diffusion coefficients have not
yet been measured, our CMS calculations predicts the diffusion behavior.

In summary, in this work we have performed a comparison between quantum DFT 
and CMS calculations in order to obtain the diffusion properties in bcc \emph{Fe-Cr}
diluted alloys. 
In accordance with Bohr's correspondence principle, as the total number of atoms
is increased, the diffusion coefficients obtained with quantum DFT calculations, 
approach the classical ones obtained with CMS. 
For DFT calculations, the electronic quantum coherence plays a crucial role that is related with the size of the simulation cell.
Also, thermal contact with the environment as the effect of killing coherence effects
making the classical behavior to emerge.
Indeed, results obtained with CMS calculations are
in good agreement with available experimental data for both solute
and solvent diffusion coefficients.  

Hence, the atomic diffusion process in metals is a classical 
phenomena, for which the large number of atoms and the temperature has
suppressed any quantum coherent effect. Then, if reliable semi empirical potentials are available,
a classical treatment of the atomic transport in metals is much convenient than DFT.


The comparison between DFT and CMS calculation is then purposed 
as a tool to investigate the effective size of quantum effects,
such as coherence length.
\vspace{-0.5cm}
\section*{Acknowledgements}
\vspace{-0.5cm}
This work was partially financed by CONICET PIP-00965/2010.

\end{document}